\documentclass[fleqn,twoside]{article}
\usepackage{espcrc2}

\usepackage{graphicx}
\usepackage[figuresright]{rotating}

\newcommand{\AmS}{{\protect\the\textfont2
  A\kern-.1667em\lower.5ex\hbox{M}\kern-.125emS}}

\def\lsim{\raise0.3ex\hbox{$\;<$\kern-0.75em\raise-1.1ex\hbox{$\sim\;$}}}
\def\gsim{\raise0.3ex\hbox{$\;>$\kern-0.75em\raise-1.1ex\hbox{$\sim\;$}}}
\def\barr{\begin{eqnarray}}
\def\earr{\end{eqnarray}}
\def\beq{\begin{equation}}
\def\eeq{\end{equation}}
\def\dmsq{\Delta m^2}
 
\def\nue{{\nu_e}}
\def\nuebar{{\bar{\nu}_e}}
\def\nux{{\nu_x}}

\def\tm{\theta_m}

\title{Supernova neutrinos: production, propagation and oscillations
\thanks{Talk given at Neutrino 2004, Paris}
}

\author{Amol Dighe\address[TIFR]{Tata Institute of Fundamental
Research, \\
Homi Bhabha Road, Mumbai 400005, India}
}

\begin{document}

\begin{abstract}
I shall review some of the recent results concerning 
the astrophysics of a core collapse supernova (SN)
and neutrino oscillations.
Neutrinos play an important role in the SN explosion,
and they also carry most of the energy of the collapse.
The energy spectra of neutrinos and antineutrinos arriving at the 
Earth incorporate information on the primary neutrino fluxes as well 
as the neutrino mixing scenario.
The analysis of neutrino propagation through the matter of the 
supernova and the Earth,
combined with the observation of a neutrino burst from 
a galactic SN, enables us to
put limits on the mixing angle $\theta_{13}$ and identify
whether the mass hierarchy is normal or inverted.
The neutrino burst also acts as an early warning signal for
the optical observation, and in addition allows us to 
have a peek at the shock wave
while still inside the SN mantle.

\vspace{1pc}
\end{abstract}

\maketitle

\section{Introduction}
\label{intro}

Neutrinos are crucial to the life and afterlife of a SN.
The current understanding of the SN explosion mechanism
suggests that neutrinos are responsible for reviving the 
stalled shock and causing the eventual explosion
\cite{bethe,book,buras}.
The protoneutron star cools through the emission of
neutrinos, which account for nearly 99\% of the
gravitational binding energy of the collapse.
The observation of the neutrino burst from a galactic SN
would shed light on many of the outstanding questions
in neutrino oscillation physics and astrophysics.

Neutrinos undergo flavour conversions on their way out 
through the mantle and envelope of the star, 
through the interstellar space, and possibly even through 
some part of the Earth before arriving at the detector.
The spectra of these neutrinos carry information about the
two mass squared differences and the $\nu_e$ flavour component 
in the three mass eigenstates.
Of course this information comes convoluted with the primary
fluxes of the neutrinos produced inside the star, and
the extraction of the mixing parameters depends crucially
on our understanding of these primary fluxes.

Recent simulations \cite{noon} indicate that 
the mean energies and relative fluxes of neutrino species
are significantly different from the traditionally 
accepted values.
Even these have large uncertainties, so that
only a few of the robust features of these spectra can be used
with confidence to extract the mixing parameters.
In spite of this limitation, it has been argued
\cite{ds,ls1}
that the observations of the $\nue$ and $\nuebar$ spectra at
the detectors on Earth may reveal the type of the neutrino
mass hierarchy and limit the value of $\theta_{13}$.
In addition, significant modifications of neutrino spectra 
that can take place if the neutrinos travel through the Earth 
matter before reaching the detector can provide 
concrete signatures for some of the neutrino mixing scenarios
\cite{cairo,ls2,sato}.

Since neutrinos are expected to arrive hours before the optical
signal from the SN, the neutrino burst serves as an early warning
\cite{snews}. 
At a water Cherenkov detector the size of SuperKamiokande (SK), 
the burst can also be used to locate the SN to within a few degrees 
in the sky \cite{beacom,ando,pointing}, 
so that optical telescopes can be pointed in the
appropriate direction.

The neutrino burst also plays an important role in our understanding 
of the supernova explosion mechanism.
Since neutrinos come unscattered from deep within the sky, we
are really looking through them at deep internal regions of 
the exploding star.
The time evolution of neutrino spectra have information 
about the shock wave
propagation encoded in them, which can be extracted at least
for certain neutrino mixing scenarios
\cite{fuller,lisi,revshock}.

This article is organized as follows. 
Sec.~\ref{prod} discusses the neutrino emission during the core
collapse and cooling, the role of neutrinos in SN explosion,
and the flavour dependence of primary fluxes and spectra.
Sec.~\ref{conv} describes the neutrino flavour conversions 
inside the star and the Earth.
Sec.~\ref{identify} discusses the extraction of neutrino mixing
parameters from the observed neutrino spectra, pointing out
how Earth matter effects can help in identifying mixing scenarios 
independently of the uncertainties in the initial fluxes.
Sec.~\ref{astro} describes how accurately the neutrino burst can 
point to the SN in advance and how features of the shock wave
can be observed through neutrinos.
Sec.~\ref{concl} concludes.


\section{Neutrino production and emission}
\label{prod}

Neutrinos and antineutrinos of all species are produced inside the
SN through pair production processes.
In addition, $\nue$ is also produced by electron capture on
protons: $p e^- \to n \nue$. 
At densities of $\rho \gsim 10^{10}$g/cc, the mean free path 
of neutrinos is much smaller than the size of the core,
so that the neutrinos are not able to stream out freely from
the core.
Even before the collapse, neutrinos of all species are trapped 
inside their respective ``neutrinospheres'' 
around $\rho \sim 10^{10}$g/cc.

When the iron core reaches a mass close to its Chandrasekhar limit, 
it becomes gravitationally unstable and collapses. 
A hydrodynamic shock is formed when the matter reaches nuclear
density and becomes incompressible. 
The shock wave dissociates the nuclei on its way outwards towards
the surface of the star.
This increases the number of protons available and consequently, 
the rate of electron capture and $\nue$ production.
As a result, when the shock wave passes through the 
$\nue$ neutrinosphere, a short $\nue$ 
``neutronization'' burst is emitted, which
lasts for $\sim$10 ms.

The object below the shock wave, the ``protoneutron star,'' then
cools down with the emission of neutrinos of all species. 
This emission takes place over a time period of $t \sim 10$ s.
The first 0.5--1 s correspond to the ``accretion phase,''
during which the matter keeps on accreting over the inner core,
emitting most of its gravitational energy in neutrinos.
Later the protoneutron star slowly contracts, cools and
deleptonizes during the so-called 
``Kelvin -- Helmholtz cooling phase'' \cite{book}.
The neutrinos emitted during these 10 s exit the star 
much before the shock wave blows the envelope up,
so the neutrinos arrive at the Earth a few hours before the
optical signal.


\subsection{Role of neutrinos in explosion}
\label{expl}

The original 
shock wave is not able to cause a SN explosion.
It loses energy in disintegrating iron nuclei, and the 
increased $\nue$ emission during the neutronization burst 
dampens the shock.
However, as more stellar matter falls onto the collapsed inner core, 
the shock is pushed to higher radii and the density and 
temperature behind the shock decrease.
At the same time, the central core begins to settle and
heats up, thus radiating more energetic neutrinos.
This results in $\sim$10\% of $\nue$ and $\nuebar$ getting 
absorbed by free neutrons and protons behind the shock.
The neutrino energy is then transferred to the shock, and if
this energy deposition is efficient enough, 
the stalled shock can be revived and drives a ``delayed''
explosion \cite{bethe}.

The ``neutrino heating'' is thus crucial for the SN explosion.
However, it is found that the energy transfer behind the
stalled shock is not efficient enough to produce explosions.
There have been no successfully simulated spherically
symmetric (1D) explosions that take into account the
elaborate transport description \cite{mezza},
and even the addition of convection in the 2D simulations
performed with a Boltzmann solver for the neutrino transport
fails to cause explosion \cite{buras}.
This suggests that either there is some missing physics 
related to the nuclear equation of state and weak interactions
in the subnuclear regime, or
there is a more fundamental problem with the neutrino driven
explosion mechanism.
(See~\cite{buras} for more details.)


\subsection{Primary neutrino fluxes and spectra}
\label{primary}

A SN core acts essentially like a neutrino blackbody source, 
but small flavour-dependent differences of the fluxes and 
spectra remain.
Since these differences are very small between 
$\nu_\mu, \nu_\tau, \bar\nu_\mu$ and $\bar\nu_\tau$, 
all these species may be represented by $\nux$.
We denote the fluxes of $\nue, \nuebar$ and $\nux$ at the 
Earth that would be observable in the absence of oscillations by 
$F_\nue^0, F_\nuebar^0$ and $F_\nux^0$ respectively. 
The energy spectra of all these ``primary'' fluxes
may be parametrized by the form ~\cite{keil1}
\begin{equation}\label{eq:spectralform}
F(E)=
\frac{\Phi_0}{E_0}\,\frac{\beta^\beta}{\Gamma(\beta)}
\left(\frac{E}{E_0}\right)^{\beta-1}
\exp\left(-\beta\frac{E}{E_0}\right)\,,
\end{equation}
where $E_0$ is the average energy, $\beta$ a parameter that typically
takes on values 3.5--6 depending on the flavour and the phase of
neutrino emission, and $\Phi_0$ the overall flux at the detector.
The values of the total flux $\Phi_0$
and the spectral parameters $\beta$ and $E_0$ are different for
$\nue, \nuebar$ and $\nux$, and are in general time dependent.
These are determined through the transport of neutrinos inside
the core and mantle of the SN.

The transport of $\nu_e$ and $\bar\nu_e$ 
inside the star is dominated by $\nu_e
n\leftrightarrow p e^-$ and $\bar\nu_e p\leftrightarrow n e^+$,
reactions that freeze out at the energy-dependent ``neutrino sphere.''
The flux and spectrum is essentially determined by the temperature and
geometric size of this emission region.  The neutron density
is larger than that of protons, so that the $\bar\nu_e$ sphere is
deeper than the $\nu_e$ sphere, explaining $\langle
E_{\nu_e}\rangle<\langle E_{\bar\nu_e}\rangle$.

For $\nux$, in contrast, the flux and spectra formation is a
three-step process. The dominating source of neutrino
scattering is the neutral-current 
nucleon scattering $\nux N\to N \nux$.  
Deep in the star thermal
equilibrium is maintained by nucleon bremsstrahlung $N
N\leftrightarrow N N \nux \bar{\nu}_x$, pair annihilation $e^-
e^+\leftrightarrow \nux \bar{\nu}_x$ and $\nu_e
\bar{\nu}_e\leftrightarrow \nux \bar{\nu}_x$, and scattering on
electrons $\nux e^-\to e^- \nux$.  The freeze-out sphere of the
pair reactions defines the ``number sphere,'' that of the
energy-changing reactions the ``energy sphere,'' and finally that of
nucleon scattering the ``transport sphere'' beyond which neutrinos
stream freely. 
                                                                                
Until recently all simulations simplified the treatment of $\nux$
transport in that energy-exchange was not permitted in $\nu
N$-scattering, $e^-e^+$ annihilation was the only pair process, and
$\nux e$-scattering was the only energy-exchange process. However,
it has been found that nucleon recoils are important for energy
exchange, 
nucleon bremsstrahlung is an important pair process, 
and $\nu_e \bar\nu_e\to \nux \bar{\nu}_x$ is 
far more important than $e^-e^+\to\nux \bar{\nu}_x$ 
as a $\nux\bar{\nu}_x$ source reaction
\cite{keil1,Raffelt:ai,Buras:2002wt}.
As a result, the recent predictions for $\nux$ fluxes differ
significantly from the traditionally used ones.

The model dependence of the fluxes is evidenced by the comparison of
typical values of the parameters in two models as shown in 
Table~\ref{models}. 
The first
is motivated by the recent Garching calculation \cite{noon} that
includes all relevant neutrino interaction rates, including nucleon
bremsstrahlung, neutrino pair processes, weak magnetism, nucleon
recoils and nuclear correlation effects.  The second is the result
from the Livermore simulation \cite{livermore} that represents
traditional predictions for flavour-dependent SN neutrino spectra that
have been used in many analyses. 
Both the models agree on 
$ {\langle E_0(\nue) \rangle} \approx 12$ MeV and 
${\langle E_0(\nuebar) \rangle} \approx 15$ MeV, 
and have consistent $\beta$ values,
but they differ
widely on ${\langle E_0(\nux) \rangle}$ and the ratios of
fluxes. 
In particular, the equipartition of energy 
assumed in the Livermore model 
is not a feature of the Garching model.

\begin{table}
\caption{Model dependence of primary fluxes
\label{models}}
\begin{tabular}{lccc}
\hline
{ Model} & ${\langle E_0(\nux) \rangle}$ &
$\frac{\Phi_0(\nue)}{\Phi_0(\nux)}$ &
 $\frac{\Phi_0(\nuebar)}{\Phi_0(\nux)}$\\
\hline
{ Garching } &
{  18} & {  0.8 }& {  0.8} \\
{ Livermore} &
{  24 }& {  2.0}&{  1.6} \\
\hline
\end{tabular}
\end{table}

In the light of the model dependence, it is important to make sure 
that the inferences drawn from the observed neutrino spectra 
do not depend strongly on the exact model parameters.

\section{Flavour conversions in matter}
\label{conv}

\subsection{Resonant conversions inside the star}
\label{resonances}

Neutrinos, while freestreaming out of the core, encounter matter
with densities ranging from $10^{10}$g/cc to almost zero.
Matter effects on the neutrino mixing, and hence on the 
flavour conversions, are crucial. Indeed,
the flavour conversions take place mainly in the resonance layers, 
where
$\rho_{\rm res} \approx m_{\rm N} \dmsq_{i}\cos 2\theta /
(2 \sqrt{2} G_{\rm F} Y_{\rm e} E)$. Here $\dmsq_{i}$ and $\theta$ 
are the
relevant mass squared difference and mixing angle of the neutrinos,
$m_{\rm N}$ is the nucleon mass,
$G_{\rm F}$ the Fermi constant and $Y_{\rm e}$ the electron fraction.
In contrast to the solar case, SN neutrinos must pass through
two resonance layers: the H-resonance layer at
$\rho_{\rm H}\sim 10^3$ g/cc corresponds to $\Delta m^2_{\rm atm}$,
whereas the L-resonance layer at
$\rho_{\rm L}\sim 10$ g/cc corresponds to $\Delta m^2_{\odot}$.
This hierarchy of the resonance densities, along with their
relatively small widths, allows the transitions in the two
resonance layers to be considered independently~\cite{ds}.

When neutrino mixing is taken into account, the 
$\nue$ and $\nuebar$ fluxes arriving at
a detector are
\begin{eqnarray}
F_{\nue} & = & p F_{\nue}^0 + (1-p) F_{\nux}^0 ~, \\ 
F_{\nuebar} & =  &\bar{p} F_{\nuebar}^0 + (1-\bar{p}) F_{\nu_x}^0 ~,
\label{feDbar}
\end{eqnarray}
where $p$ and $\bar{p}$ are the
survival probablilities of $\nue$  and $\nuebar$ respectively.

The neutrino survival probabilities can be characterized by the degree
of adiabaticity of the resonances traversed, which are directly
connected to the neutrino mixing scheme.
In particular, whereas the L-resonance is always adiabatic and
appears only in the neutrino channel, the adiabaticity of
the H-resonance depends on the value of $\theta_{13}$, and
the resonance shows up in the neutrino or antineutrino channel
for a normal or inverted mass hierarchy respectively.
Table~\ref{tab-pbar} shows the survival probabilities 
in various mixing scenarios.
For intermediate values of $\theta_{13}$, i.e.
$10^{-5}\lsim\sin^2 \theta_{13} \lsim 10^{-3}$,
the survival probabilities depend on energy as well
as the details of the density profile of the SN.

\begin{table}
\caption{Survival probabilities for neutrinos, $p$, and
antineutrinos, $\bar{p}$, in various mixing scenarios 
\label{tab-pbar}}
\begin{tabular}{llccc}
\hline
 &
Hierarchy &  $\sin^2 \theta_{13}$  &  $p$ &  $\bar{p}$ \\
\hline
A & Normal & $\gsim 10^{-3}$  & 0  & $\cos^2\theta_\odot$ \\
B & Inverted &  $\gsim 10^{-3}$ &  $\sin^2\theta_\odot$ &  0 \\
C & Any & $\lsim 10^{-5}$  & $\sin^2\theta_\odot$
&  $\cos^2\theta_\odot$ \\
\hline
\end{tabular}
\end{table}

Scenarios A, B and C are the ones that can in principle be 
distinguished through the observation of a SN neutrino burst.

\subsection{Oscillations inside the Earth matter}
\label{earth}

If the neutrinos travel through the Earth before reaching 
the detector, the neutrinos undergo oscillations inside 
the Earth and the survival probabilities change.
This change however occurs only in those scenarios 
in Table~\ref{tab-pbar} where the value of the survival
probability is nonzero. The expressions in this section
are to be understood in that context.

For antineutrinos that pass only through the mantle with roughly
a constant density, the survival probability $\bar{p}^D$ is 
\begin{equation}
\bar{p}^D  \approx  \cos^2 \theta_{12}
+ \bar{A}_m 
~\sin^2 \left( \overline{\dmsq_m} L_m y \right)\,.
\label{pbar}
\end{equation}
where 
$\overline{\dmsq_m}$ is the mass squared difference between
$\bar{\nu}_1$ and $\bar{\nu}_2$ inside the mantle in units of
$10^{-5}$ eV$^2$, and $L_m$~is the distance traveled through the
mantle in units of 1000~km.  The ``inverse energy'' parameter 
$y$ is defined as
$y \equiv 12.5~{\rm MeV}/E$
where $E$ is the neutrino energy.
The coefficient of the oscillating term is 
$\bar{A}_m \equiv - \sin 2\tm \sin (2\tm - 2\theta_{12})$.

When neutrinos travel through both the mantle and the core, 
the sharp density jumps give rise to the survival probability
of the form
\beq
\bar{p}^D \approx \cos^2 \theta_{12} + \sum_{i=1}^7 \bar{A}_i
\sin^2(\phi_i/2) 
\label{pdcore}
\eeq
in the two-layer model of the Earth, 
where the coefficients $ \bar{A}_i$ are functions of the mixing
angle $\theta_{12}$ in vacuum, mantle and core.
The phases $\phi_i$ depend on the distance travelled through
the Earth matter and the values of $\overline{\dmsq}$
in the mantle and the core \cite{corewiggles}.


\section{Distinguishing between neutrino mixing scenarios}
\label{identify}

The only SN observed in neutrinos till now, SN1987A,
yielded only $\sim$20 events. Though it confirmed our
understanding of the SN cooling mechanism, the number
of events was too small to say anything concrete
about neutrino mixing
(see \cite{ls-sn87} and references therein).
On the other hand, if a SN explodes 
in our galaxy at 10 kpc from the Earth,
we expect $\sim$10000 events at SK. 
With the mixing scenarios A, B and C having clearly distinct
survival probabilities for $\nue$ and $\nuebar$,
the task of distinguishing between the scenarios 
naively seems straightforward: measure the neutrino fluxes 
arriving at the Earth, and determine the
values of $p$ and $\bar{p}$.

There are a few major problems, though. 
With the current detectors, one can obtain a 
statistically significant and clean spectrum 
only of $\nuebar$, through the inverse beta reaction
$\nuebar p \to n e^+$ at a water Cherenkov or scintillation 
detector.
It is possible to obtain a clean $\nue$ spectrum at a 
heavy water detector like SNO through 
$\nue d \to p~ p ~e^-$, or 
at a liquid Ar detector through
$\nu_e + ~^{40}{\rm Ar} \to ~^{40}{\rm K}^* + e^-$,
but the sizes of the current detectors of these kinds,
and hence the number of events expected in them, are 
very small. A large liquid Ar detector, as suggested in
\cite{botella1}, would be very significant,
though technologically challenging, in this context.

Secondly, the primary spectra are poorly known. 
The uncertainties in the values of $E_0$ for $\nue, \nuebar$
and $\nux$ mean that by merely observing a mixed spectrum,
it is not possible to determine the extent of $\nux$
component in it.
A number of observables have been suggested 
\cite{ds,ls1,botella2}
that can distinguish between different scenarios if
the primary spectra obey certain form or if the parameters
lie within some bounds, 
but it has turned out to be very difficult to come up with 
clean observables that do the job independently of
any assumption about the primary spectra.

The presence or absence of Earth effects, however, 
can be exploited to 
detect model independent signatures of mixing scenarios.
Earth effects manifest themselves in two ways.
Firstly, the total number of events and the spectral shape
changes, this can be checked by comparing the neutrino
signal at two or more detectors such that the neurinos
travel different distances through the Earth before 
reaching them.
Secondly, Earth effect oscillations are introduced,
which may be identified even at a single detector.
These two approaches will be illustrated in the next two 
subsections.


\subsection{Comparing signals at multiple detectors}
\label{multiple}

At least one of the existing detectors (SK, SNO or LVD) should
observe the SN neutrinos through the Earth for 
the location of the SN in a large fraction (60\%) of the sky
\cite{ls2}.
However, for a SN at 10 kpc one can only get a statistical 
significance of 2--3$\sigma$. 
In order to get a larger significance, at least two detectors 
of the size of SK or larger are needed \cite{ls2}.

In this context, the km$^3$ ice Cherenkov detector in Antarctica, 
IceCube, can be used as a co-detector with SK or its larger
version, HyperKamiokande (HK).
IceCube is primarily meant for detecting individual neutrinos 
with energy $\gsim 150$ GeV.
However, during a galactic SN burst, the number of Cherenkov
photons detected by the optical modules would increase
much beyond the background fluctuations, so that the burst
as a whole can be identified \cite{halzen}.
It is not possible to detect inividual neutrinos and measure 
their energies, but it is possible to measure the time dependence 
of luminosity which is proportional
to the total number of Cherenkov photons detected.
Indeed, for a SN burst at 10 kpc, the luminosity can be 
determined to a statistical accuracy of $\sim 0.25$\% \cite{icecube}.

The Earth effects can change the luminosity by upto 10\%.
So if the neutrinos travel different distances through the Earth
before reaching SK/HK and IceCube, the ratio of luminosities 
at the two detectors can show evidence for the Earth effects.
Moreover, for typical numerical SN simulations, 
the Earth effect is time dependent and most notably differs 
between the early accretion
phase and the subsequent Kelvin--Helmholtz cooling phase
by 3--4\%. 
This indicates that there is no need even of the absolute
calibration of either of the detectors, 
one just has to search for a temporal variation of the 
relative detector signals of a few percent. 
The large number of optical modules in
IceCube renders this task statistically possible
\cite{icecube}.
The accuracy of luminosity measurement in SK/HK
would be the limiting factor.

The relative locations of SK and IceCube imply that 
for the SN in a large portion of the sky, 
it is observed by only one of the detectors through the Earth.
This makes the SK/HK--IceCube comparison 
an interesting prospect.


\subsection{Identifying Earth effects at a single detector}
\label{single}

The Earth matter effects on supernova
neutrinos traversing the Earth mantle give rise to a specific
frequency in the ``inverse energy'' spectrum of these neutrinos,
as can be seen by writing
the net $\nuebar$ flux at the detector 
using eqs.~(\ref{feDbar}) and (\ref{pbar}) in the form
\barr
F_{\nuebar}^D & = &  \sin^2 \theta_{12}  F_\nux^0 +
\cos^2 \theta_{12} F_{\nuebar}^0 + \nonumber \\
& & \phantom{\sin^2 \theta_{12}  F_\nux^0} \Delta F^0
\bar{A}_m \sin^2(k_m y /2) \,,
\label{feDbar-y}
\earr
where $\Delta F^0 \equiv (F_{\nuebar}^0 - F_\nux^0)$ depends
only on the primary neutrino spectra, and
$k_m \equiv 2 \overline{\dmsq_m} L_m$.
Note that $\bar{A}_m$ 
depends only on the mixing parameters and is independent of the
primary spectra.

The last term in Eq.~(\ref{feDbar-y}) is the Earth oscillation term
that contains a frequency $k_m$ in
$y$, the coefficient $\Delta F^0 \bar{A}_m$ being a relatively slowly
varying function of $y$.  The first two terms in Eq.~(\ref{feDbar-y})
are also slowly varying functions of $y$, and hence contain
frequencies in $y$ that are much smaller than $k_m$.  The dominating
frequency $k_m$ is the one that appears in the modulation of the
inverse-energy spectrum.
                                                                                
The frequency $k_m$ is completely independent of the primary
neutrino spectra, and indeed can be determined to a good accuracy from
the knowledge of the solar oscillation parameters, the Earth matter
density, and the position of the SN in the sky.
Therefore, Earth effects can be identified merely by
identifying the presence of this oscillation frequency in the
observed spectrum.
This may be achieved by taking a Fourier transform of the
inverse-energy spectrum and looking for peaks in the power spectrum
\beq
G_N(k) =  \frac{1}{N} \left| \sum_{\rm events} 
e^{i k y_{\rm event}} 
\right| ^2 ~.
\eeq
The peak corresponding to the oscillation frequency $k_m$
emerges on top of the random background fluctuations.
The position of this peak is insensitive to the primary
spectra \cite{fourier}.

If both the mantle and the core are crossed before the neutrinos reach
the detector, as many as seven distinct frequencies are present in the
inverse energy spectrum, as can be see from eq.~(\ref{pdcore}).
 However only three peaks are dominant
in the  power spectrum due to the hierarchy in the
$\bar{A}_i$ values. The increase in the number of
expected peaks leads to an easier identification of the Earth matter
effects \cite{corewiggles}.

The energy resolution of the detector turns out to be crucial 
in detecting the Earth effect oscillations, since bad energy
resolution tends to smear out the modulations in the 
energy spectrum.
The comparison between a simulated megaton water Cherenkov
detector and a 32 kt scintillation detector \cite{corewiggles}
shows that 
the better resolution of the scintillator detector almost 
compensates for the much larger water Cherenkov detector size.  
On the other hand, the
worse energy resolution in water Cherenkov detectors does not only
imply the need of a larger volume but it also suppresses significantly
the peaks at higher frequencies, in contrast to the case of
scintillator detectors.

Only scenarios A and C allow observable Earth effects in $\nuebar$.
Therefore, the observation of a Fourier peak in $\nuebar$ 
eliminates scenario B independent of SN models. Similarly,
if earth effects are observed in the $\nue$ spectrum,
scenario A may be eliminated.


\section{Neutrinos for SN astrophysics}
\label{astro}

\subsection{Pointing to the SN in advance}
\label{pointing}

Determining the accuracy to which SN can be located in the sky with
neutrinos alone is important for two reasons.
Firstly, the neutrino burst precedes the optical explosion by
several hours so that an early warning can be issued to the astronomical
community~\cite{snews}, specifying the direction to look for
the explosion.
Secondly, in the absence of any SN observation in the electromagnetic
spectrum, a reasonably accurate location in the sky is crucial for
determining the neutrino Earth-crossing path to various detectors
since the Earth matter effects on SN neutrino oscillations may well
hold the key to identifying the neutrino mixing scenario.

The best way to locate a SN by its core-collapse
neutrinos is through the directionality of 
the elastic scattering $\nu e^- \to \nu e^-$ events
in a water Cherenkov detector
such as SK~\cite{beacom,ando}.
The directionality of this reaction is primarily limited 
by the angular resolution of
the detector and to a lesser degree by the kinematical deviation of
the final-state electron direction from the initial neutrino.

The pointing accuracy 
is further strongly degraded by the
inverse beta reactions $\bar\nu_e p\to n e^+$ that are nearly
isotropic and about 30--40 times more frequent than the elastic
scattering events.  Recently it was proposed to add to the water a
small amount of gadolinium, an efficient neutron absorber, that would
allow one to detect the neutrons and thus to tag the inverse beta
reactions~\cite{gadzooks}.  
Removing this major background would still leave one with 
the nearly isotropic oxygen reaction 
$\nu_e+{}^{16}{\rm O}\to{\rm X}+e^-$.
No clean separation of this background is possible.

The pointing accuracy also has a weak dependence on the neutrino
mixing scenario.
It has been found \cite{pointing} that, 
for the ``worst case'' mixing scenario and for
the tagging efficiency $\epsilon_{\rm tag}=0$, 
at 95\% C.L.\ the pointing accuracy at SK is
$7.8^\circ$, which improves to $3.6^\circ$ 
for $\epsilon_{\rm tag} =80\%$
and $3^\circ$ for $\epsilon_{\rm tag} = 1$.
Thus, neutron tagging results in
nearly a factor of 3 improvement in the pointing angle,
which corresponds to almost an order of magnitude improvement 
in the area of the sky in which the SN is located.


\subsection{Tracking the shock wave in neutrinos}
\label{shock}

The passage of the shock wave through the density of the 
H-resonance ($\rho \sim 10^3$ g/cc) a
few seconds after the core bounce may
break adiabaticity, thereby modifying the spectral
features of the observable neutrino flux. Therefore, it is conceivable
that a neutrino detector can measure the modulation
of the neutrino signal caused by the shock-wave propagation, an effect
first discussed by Schirato and Fuller in a seminal
paper~\cite{fuller} and elaborated by a number of subsequent
authors~\cite{ls1,lisi,takahashi}.
Since the density of the
H-resonance depends on energy, the observation of such a modulation
in different neutrino energies would allow one to trace the shock
propagation. On the other hand, the occurrence of this effect
depends on the sign of $\Delta m^2_{31}$ and the value of
$\theta_{13}$, so that observing it in the $\bar\nu_e$ spectra, the
experimentally most accessible channel, would
imply that the neutrino mass ordering is inverted and that
$\sin^2\theta_{13}\gg 10^{-5}$.
This corresponds to the elimination of scenarios A and C.

Some time after the onset of the explosion
a neutrino-driven baryonic wind develops and collides with the
earlier, more slowly expanding supernova ejecta. 
This gives rise to a ``reverse shock''. This is a generic
feature of all SN simulations, although
the exact propagation history depends on the detailed
dynamics during the early stages of the supernova explosion.
The simultaneous propagation of a direct and a reverse shock wave 
manifests itself in a ``double dip'' feature in the 
time evolution of observables like the average neutrino energy
and the number of events \cite{revshock}. 

If the time evolution of the number of events
in different energy bins is observed, the 
positions of the two dips in time can be connected to the
positions of the forward and reverse shock. 
Indeed, the number of shock waves present in a region with 
any given density $\sim 10^3$~g/cc can
be extracted from the data by considering the time evolution of
the number of events in the energy bin corresponding to that 
resonant density \cite{revshock}.
An extrapolation would allow one to trace the positions of the forward
and the reverse shock waves for times between 1--10~s.
Since the neutrino conversion probabilities are energy dependent
during the passage of the shocks through the H-resonance, neutrino
oscillations can be detected even if the energy spectra of different
neutrino flavours have the same shape but different luminosities.


\section{Hoping for a catastrophe}
\label{concl}

The observation of neutrinos from a core collapse SN 
is expected to reap a rich scientific harvest.
It will immensely improve our understanding of SN
astrophysics.
If the value of $\theta_{13}$ and the type of neutrino
mass hierarchy is already determined at terrestrial experiments,
concrete information on the primary neutrino fluxes will be
obtained. 
On the other hand, if the burst takes place before the
mixing parameters are measured, the limits obtained 
can guide us in deciding on the design parameters 
of future long baseline experiments.

A galactic SN burst is a rare phenomenon, expected to occur 
only 2--3 times in a century. 
It is therefore imperative that we are ready with suitable 
long term detectors that will observe the relevant signals.
In the meanwhile, better theoretical understanding of 
neutrino transport inside the SN,
combined with more accurate measurements of the neutrino mixing 
parameters, will equip us for 
making the most of the cosmic catastrophe.

I thank the organisers of Neurino 2004 for their hospitality.

\end{document}